\documentclass[
    ,final            
  ]
  {aipproc}

\layoutstyle{6x9}

\begin{document}

\title{Fusion at deep subbarrier energies: \\
potential inversion revisited}

\classification{
25.70.Jj,24.10.Eq,03.65.Sq,03.65.Xp}

\keywords      {Heavy-ion fusion, coupled-channels method, barrier distribution, inter-nucleus potential}

\author{K. Hagino}{
  address={Department of Physics, Tohoku University, Sendai 980-8578, Japan}
}

\author{N. Rowley}{
  address={Institut Pluridisciplinaire Hubert Curien (UMR 7178: CNRS/ULP), 
23 rue du Loess, F-67037 Strasbourg Cedex 2, France}
}

\begin{abstract}
For a single potential barrier, 
the barrier penetrability can be inverted 
based on the WKB approximation to yield the barrier thickness.  
We apply this method to 
heavy-ion fusion reactions at energies
well below the Coulomb barrier and directly determine the inter-nucleus
potential between the colliding nuclei. 
To this end, we assume 
that fusion cross sections at deep subbarrier energies are
governed by the lowest barrier in the barrier distribution. 
The inverted 
inter-nucleus potentials 
for the $^{16}$O +$^{144}$Sm and 
$^{16}$O +$^{208}$Pb reactions show that 
they are much thicker than phenomenological
potentials. We discuss a consequence of such thick potential 
by fitting the inverted potentials with the Bass function. 
\end{abstract}

\maketitle


\section{Introduction}

Nuclear reactions are primarily governed by 
the nucleus-nucleus potential. 
Several methods have been 
proposed to compute the real part of the inter-nuclear potential. 
Among them, the double folding model has been 
often employed and has enjoyed a success 
in describing elastic and inelastic scattering for many systems
\cite{SL79,BS97,KS00}. 
The Woods-Saxon form, which fits the double folding potential in 
the tail region, 
has also often been used to parametrize the
inter-nuclear potential \cite{BW91}. 

In recent years, many experimental evidences have accumulated that
show that the double folding potential fails to account for
the {\it fusion} cross sections at energies close to the Coulomb 
barrier \cite{L95,NBD04,NMD01,HDG02,HRD03,GHDN04,DHNH04}. 
This trend has become even more apparent 
in the recent measurements of fusion cross 
sections at extreme subbarier energies, that show a much 
steeper fusion excitation functions as compared with theoretical
predictions \cite{Jiang}. 

The scattering process 
is sensitive mainly to the surface region of the nuclear potential, 
while the fusion reaction is also relatively sensitive to the inner part. 
The double folding potential and the Woods-Saxon potential are 
reasonable in the surface region \cite{WHD06}. 
However, 
it is not obvious whether they 
provide reasonable parametrizations 
inside the Coulomb barrier, where the 
colliding nuclei significantly overlap with each other
\cite{NBD04,DP03,ME06}. 

In this contribution, we discuss
the radial shape of the inter-nucleus potential 
inside the Coulomb barrier and 
investigate its deviation from the conventional parametrizations. 
To this end, 
we shall first determine 
the inter-nuclear potential directly from the
experimental data without assuming any parametrization \cite{HW07}. 

\section{Inter-nucleus potential from fusion data}

There have been lots of attempts to determine an inter-nucleus potential 
directly from experimental fusion excitation functions \cite{VAL81}. 
In the 70's, it was fashionable to plot a fusion excitation function 
as a function of $1/E$\cite{SGBM76}. Since the classical fusion cross 
section is given by $\sigma(E)=\pi R_b^2 (1-V_b/E)$, where $R_b$ and $V_b$ 
are the position and the height of the Coulomb barrier, respectively, 
the value of $R_b$ and $V_b$ can be read off from such plot (see also 
Ref. \cite{RKL89}). Bass analysed the first derivative of $E\sigma$, 
that is, $d(E\sigma)/dE$, and extracted an empirical inter-nucleus 
potential \cite{B77}. He fitted the deduced potential using a function 
\begin{equation}
V(r)\propto \frac{1}{A\exp[(r-R_P-R_T)/d_1]+B\exp[(r-R_P-R_T)/d_2]},
\end{equation}
where $A,B, d_1$, and $d_2$ are adjustable parameters. 
This potential (the Bass potential) has been widely used, especially 
for fusion of massive systems. 
 
A more direct way to determine an inter-nucleus potential is to 
use the potential inversion method based on the WKB 
approximation \cite{CG78,BKN83}. 
For a single channel system with a potential $V(r)$, 
the inversion formula relates the thickness of the potential, 
{\it i.e.,} the distance between the two classical turning points at a
given energy $E$, with the classical action $S$ as 
\begin{equation}
t(E)\equiv r_2(E)-r_1(E) 
=-\frac{2}{\pi}\sqrt{\frac{\hbar^2}{2\mu}}
\,\int^{V_b}_EdE'\,\frac{\left(\frac{dS}{dE'}\right)}{\sqrt{E'-E}},
\end{equation}
where $\mu$ is the reduced mass between the colliding nuclei. 
The classical action $S(E)$ is given
by 
\begin{equation}
S(E)=\int^{r_2(E)}_{r_1(E)}dr\,\sqrt{
\frac{2\mu}{\hbar^2}(V(r)-E)}, 
\end{equation}
and can be obtained once the penetrability $P(E)$ is found in some
way using the WKB relation $P(E)=1/[1+e^{2S(E)}]$. 

Balantekin {\it et al.} 
assumed a one-dimensional energy independent local potential, 
and applied this method \cite{BKN83}. 
They found that the
inversion procedure leads to an
unphysical multi-valued potential for heavy systems. 
This
analysis has actually provided a clear evidence for inadequacy of
the one-dimensional barrier passing model for heavy-ion fusion reactions, 
and has triggered to develop the coupled-channels approach. 

\section{Potential inversion revisited} 

The main reason why 
Balantekin {\it et al.} obtained the unphysical inter-nucleus potentials is 
that they did not take into account the channel coupling effect, 
which has by now been well understood in terms of 
barrier distribution \cite{L95,RSS91,DHRS98,BT98,HB04}. 
We can then ask ourselves whether a well behaved potential is 
obtained if one explicitly takes into account the channel coupling 
effect. 
We address this question by applying the inversion procedure only to the lowest
barrier in the barrier distribution. 

In heavy-ion fusion reactions, 
it is well known that the $s$-wave penetrability for the Coulomb
barrier can be approximately obtained from the fusion cross section 
$\sigma_{\rm fus}$ as \cite{BKN83,RSS91,DHRS98,BT98}
\begin{equation}
P(E)=\frac{d}{dE}\,\left(\frac{E\sigma_{\rm fus}}{\pi R^2}\right). 
\label{1stderivative}
\end{equation}
In the previous application of the inversion formula by Balantekin
{\it et al.}, they assumed that the penetrability so obtained was
resulted from the penetration of a one dimensional energy independent 
potential\cite{BKN83}. 
Instead, here we assume 
that the penetrability $P$ is given as a weighted sum of contribution from 
many distributed barriers, where the distribution arises due to a coupling 
of the relative motion between the colliding nuclei to nuclear intrinsic
degrees of freedoms such as collective vibrational or rotational
excitations. In this eigen channel picture, the penetrability is given by, 
$P(E)=\sum_nw_n P_n(E)$,
where $P_n$ is the penetrability for the $n$-th eigen-barrier and 
$w_n$ is the corresponding weight factor. 

At energies below the lowest eigen barrier ({\it i.e.}, the adiabatic
barrier) in the barrier distribution, one 
expects that only the lowest barrier contributes to the total 
penetrability, $P(E)\approx w_0 P_0(E)$. 
This indicates that one can apply the inversion formula to 
the lowest eigen barrier using fusion cross sections at 
deep subbarrier energies, after correcting the weight factor. 
The height of the lowest barrier could be estimated from the 
lowest peak in the fusion barrier distribution \cite{HW07}. 

Notice that 
the inversion formula yields only the barrier thickness, $t(E)$, and
one has to supplement either the outer or the inner turning points 
to determine the radial shape of the potential \cite{BKN83}. 
We estimate the {\it outer} turning point $r_2(E)$ using the 
Coulomb interaction of point charge and the 
Woods-Saxon nuclear potential, 
%
with the range parameter of 
$R_0=\sum_{i=P, T} 
\left(1.233A_i^{1/3}-0.98A_i^{-1/3}\right)+0.29 ~~~~({\rm fm})$, 
and the diffuseness parameter of $a$=0.63 fm. 
We adjust the depth $V_0$ in order to reproduce the barrier height
$V_b$ determined from the peak position of the barrier distribution. 
Since the Coulomb term dominates at the outer turning point, 
except for the region near the barrier top, 
the inverted potential is insensitive to the actual shape of nuclear 
potential employed to estimate the outer turning point. 
The Woods-Saxon potential 
determines not only the outer
turning point but also the position of the potential barrier, $R_b$. 
In the actual application of the inversion formula shown below, we 
smooth the data points with a
fifth-order polynomial fit to the function 
$\ln[E\sigma_{\rm fus}/\pi R^2]$ \cite{BKN83}. 
We have confirmed that the results do not significantly change even if
we use a higher order polynomial fit. 
We also fit the lowest peak of the barrier distribution using the Wong
formula \cite{W73} in order to accurately estimate the barrier height $V_b$. 

\begin{figure}[htb]
\includegraphics[scale=0.5,clip]{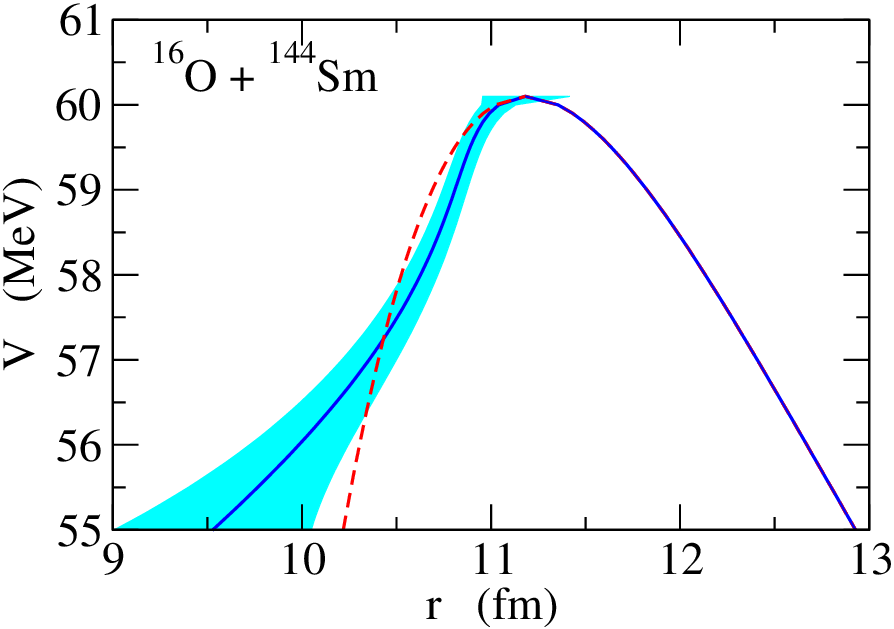}
\caption{
The adiabatic potential for the 
$^{16}$O+$^{144}$Sm reaction obtained with the inversion method. 
The dashed line is a barrier due to a phenomenological Woods-Saxon potential. } 
\end{figure}

The resultant inverted inter-nucleus potentials 
for the $^{16}$O+$^{144}$Sm 
and the $^{16}$O+$^{208}$Pb 
systems 
are shown in Figs. 1 and 2, respectively. 
We used the measured cross sections reported in Refs. \cite{L95,DHLN06} 
for the inversion procedure. 
The uncertainty of the inverted potential is estimated in the same way
as in Ref. \cite{BKN83}. 
The dashed line shows the barrier due to the Woods-Saxon 
potential 
used to
estimate the outer turning points. 
One clearly sees that the inverted potentials are much thicker than the 
phenomenological potentials at low energies, although 
they are close to the phenomenological potentials at energies close to 
the potential barrier. 
This trend is opposite to what Balantekin {\it et al.} found
in the previous analysis. If there was an unresolved peak in the
barrier distribution below the main peak, one would obtain a much
thinner barrier than the phenomenological potential, as in the previous 
analysis. 
Having thick barriers, rather than thin barriers, we are 
convinced that the main peak of the barrier distribution for 
the $^{16}$O+$^{144}$Sm and the $^{16}$O+$^{208}$Pb 
reactions indeed consist of the lowest eigen
barrier. 

\begin{figure}[htb]
\includegraphics[scale=0.5,clip]{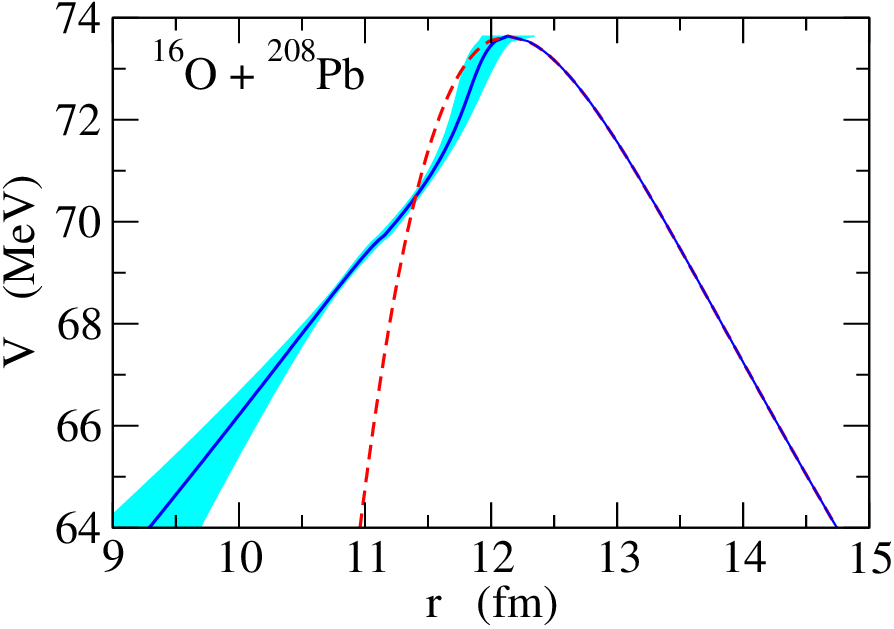}
\caption{
Same as Fig. 1, but for the 
$^{16}$O+$^{208}$Pb reaction. }
\end{figure}

The thicker the potential is, the smaller the penetrability is, and
also the stronger the energy dependence of the penetrability is. 
The thick potentials barrier obtained for the 
$^{16}$O+$^{144}$Sm and $^{16}$O+$^{208}$Pb systems are thus consistent 
with the recent experimental observations  
\cite{Jiang,DHLN06} that the fusion excitation
function is much steeper than theoretical predictions at deep
subbarrier energies. Although the present analysis does not exclude a
possibility of a shallow potential \cite{ME06}, the present study
suggests that the origin of the
steep fall-off phenomenon of fusion cross section can be 
at least partly attributed to the departure of inter-nuclear potential 
from the Woods-Saxon shape. 

For the $^{16}$O+$^{208}$Pb system shown in Fig. 2, the deviation of
the inverted potential from the phenomenological potential 
starts to occur at around $E=70.4$ MeV. 
It is amusing to notice that this energy is very close to 
the potential energy at the contact configuration estimated with the 
Krappe-Nix-Sierk potential\cite{KNS79,IHI07}. 
Inside the touching configuration, the potential 
represents 
the fission-like adiabatic potential energy surface. 
The effect of such one-body potential has been considered recently
and is shown to account well for the steep fall-off phenomena of
fusion cross sections \cite{IHI07}. 
The inverted potentials which we obtain are thus intimately related to
the one-body dynamics for deep subbarrier fusion reactions. 

\section{Discussion}

Although the inverted potentials shown in Figs. 1 and 2 are well 
behaved, there remains a question whether the inter-nucleus potential 
in itself is actually thick or it simply 
mocks up some dynamical effects such as 
energy and angular momentum dissipations. 
To address this question, we fit the inverted potential with 
the Bass function given by Eq. (1). 
A motivation to use the Bass function is that it leads to a thicker 
potential than a double folding potential. 
This is demonstrated in Fig. 3 for the $^{16}$O+$^{208}$Pb system. 
One can see that the Bass potential is much thicker than the Woods-Saxon 
potential with the surface diffuseness parameter of $a$=0.65 fm, although 
both the potentials are similar to each other in the tail region. 
Notice that the thickness of the Bass potential is similar to that of 
the Woods-Saxon potential with $a$=1.0 fm. 

\begin{figure}[htb]
\includegraphics[scale=0.5,clip]{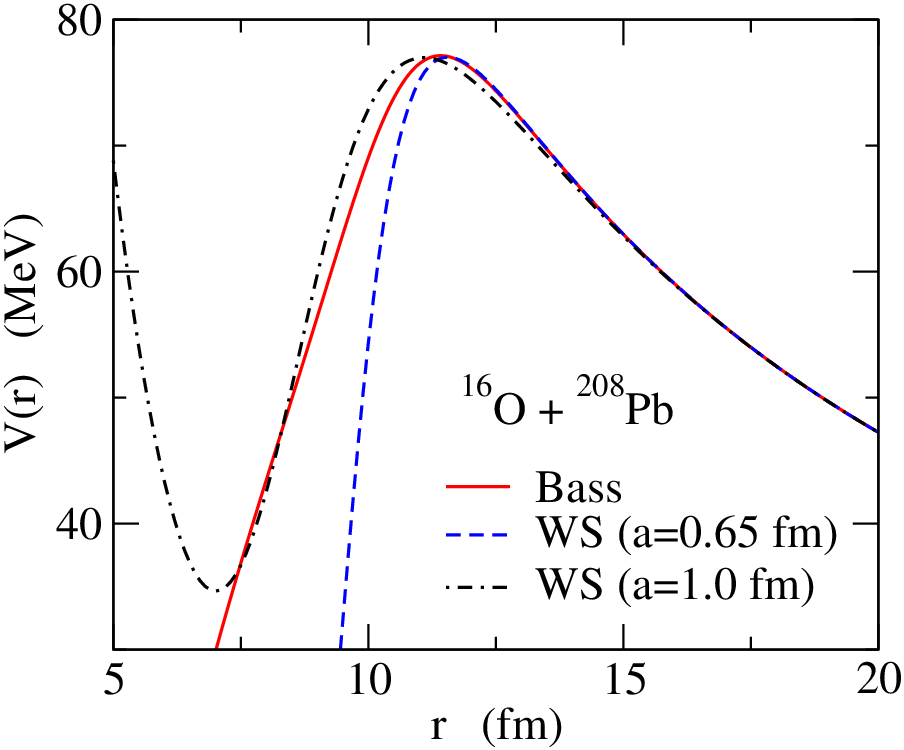}
\caption{
Comparison of the Bass potential with the Woods-Saxon potential with two 
different values of surface diffuseness parameter $a$ 
for the $^{16}$O+$^{208}$Pb system. }
\end{figure}

Figure 4 shows the result of the fitting for 
the $^{16}$O+$^{208}$Pb system. In 
the original Bass potential, the parameters $A,B,d_1$ and $d_2$ take 
the value of $A$=0.03 MeV$^{-1}$, $B=0.0061$ MeV$^{-1}$, $d_1=$3.3 fm, 
and $d_2$=0.65 fm \cite{B77}. 
We slightly change the value of $A$ to 0.05 MeV$^{-1}$, and 
adjust the depth of the potential. 
The dot-dashed line in Fig.4 shows the potential obtained in 
this way, which we use as the bare potential for the coupled-channels 
calculation. The lowest eigen-barrier obtained by diagonalizing 
the coupling matrix at each position $r$ is denoted by the dashed line. 
To this end, we include the single octupole excitation in $^{16}$O and 
the double octupole excitations in $^{208}$Pb, using the CCFULL 
scheme \cite{HRK99}. 
The resultant eigen-potential fits well the inverted potential 
obtained in the previous section. 

\begin{figure}[htb]
\includegraphics[scale=0.5,clip]{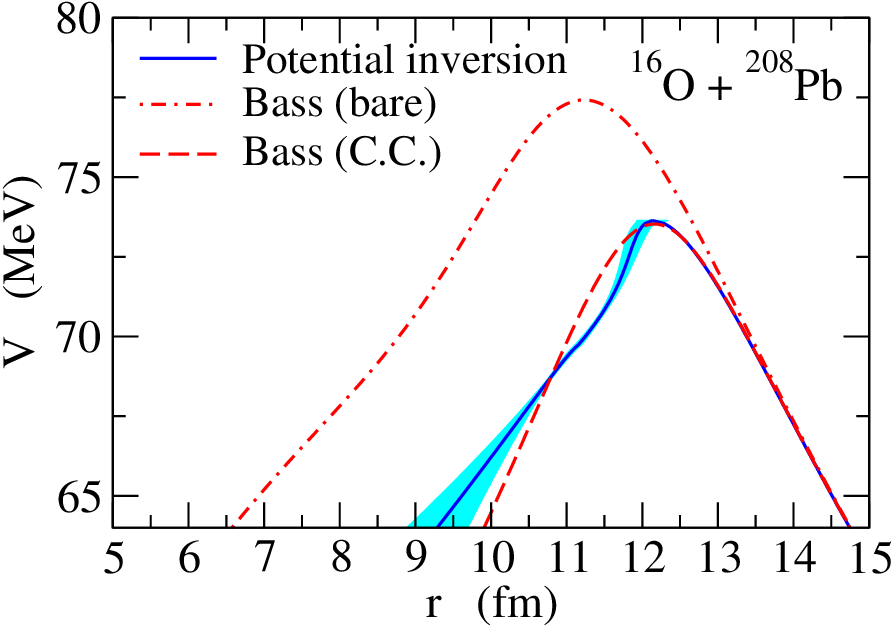}
\caption{
The result of the fitting of the inverted potential with the Bass function 
given by Eq. (1) for the $^{16}$O+$^{208}$Pb system. 
The dot-dashed line shows a bare potential, while the dashed line 
is the lowest eigen-barrier obtained by including the octupole 
excitations in the colliding nuclei. }
\end{figure}

The result of coupled-channels calculation with this potential is shown in 
Fig. 5. This calculation well reproduces the steep fall-off of fusion cross 
sections at deep subbarrier energies. A small discrepancy around 
$E_{\rm cm}\sim 70$MeV may be accounted for by including the one-neutron 
pick-up transfer channel, as was recently pointed out by Esbensen and 
Misicu \cite{EM07}. However, this calculation largely underestimates 
the fusion cross sections at energies above the Coulomb barrier. 
We have checked that it is the case even when we use an internal imaginary 
potential for fusion, instead of the incoming wave boundary condition. 
Since the potential which fits the deep subbarrier data does not reproduce 
the measured fusion cross sections at higher energies, this may be an 
indication of some other missing dynamical effect, either at deep subbarrier 
energies or at energies above the barrier (or both). 
The same conclusion has been reached in Ref. \cite{DHLN06}. 

\begin{figure}[htb]
\includegraphics[scale=0.5,clip]{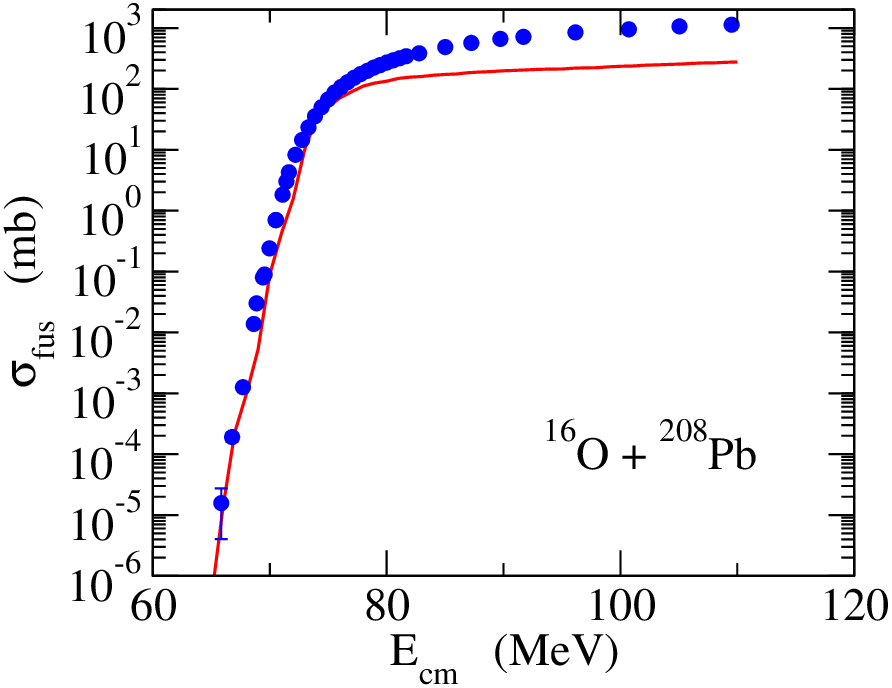}
\caption{
The fusion cross sections 
for the $^{16}$O+$^{208}$Pb system 
obtained with the coupled-channels calculations 
with the potential shown in Fig. 4.} 
\end{figure}

Another indication of a missing dynamical effect could be seen 
in the experimental barrier distributions for 
the $^{16}$O$+^{144}$Sm system. 
It has been known that the fusion barrier distribution for 
this system has a clear double-peaked structure, whereas 
the higher peak is significantly smeared in the quasi-elastic barrier distribution 
\cite{thim-95}. 
We have recently performed the coupled-channels calculations for 
this system with several coupling schemes \cite{ZH08}. 
Our calculations indicate
that, within the same coupling scheme, the quasi-elastic and fusion barrier
distributions are always similar to each other, and the difference in 
the experimental barrier distributions cannot be accounted for 
within the standard coupled-channels approach. 
This indeed suggests that some physical effects, besides 
the collective excitations in the colliding nuclei, have to be 
taken into account in order to explain the fusion and quasi-elastic scattering 
simultaneously. 

\section{Summary}

We applied the potential inversion method, which relates
the potential penetrability to the thickness of the potential barrier, 
in order to investigate the radial dependence of the inter-nucleus
potential for heavy-ion fusion reactions. 
To this end, we assumed that the tunneling is well described by the
lowest adiabatic barrier at deep subbarier energies, and extracted the
penetrability by combining the experimental barrier distribution and 
fusion cross sections. 
We found that the resultant potential for the 
$^{16}$O+$^{144}$Sm and $^{16}$O+$^{208}$Pb systems 
is much thicker than a barrier obtained with 
a phenomenological Woods-Saxon potential. 
This indicates that the steep fall-off phenomenon of fusion cross
sections recently observed in several systems can be partly accounted
for in terms of a deviation of inter-nuclear potential from the 
Woods-Saxon shape, although some dynamical effects are also expected to 
play a role. 




\begin{theacknowledgments}
This work is based on a collaboration with Y. Watanabe. 
We thank T. Ichikawa 
for useful discussions. 
This work was supported by the Grant-in-Aid for Scientific Research,
Contract No. 19740115, 
from the Japanese Ministry of Education, Culture, Sports, Science and
Technology. 
\end{theacknowledgments}



\end{document}